\documentclass[12pt,preprint]{aastex}

\newcommand{\ms}{\mbox{m s$^{-1}~$}}

\newcommand{\msun}{M$_{\odot}$}
\newcommand{\mstar}{M$_{\ast}$}

\newcommand{\mjup}{M$_{\rm JUP}~$}
\newcommand{\mjupe}{M$_{\rm JUP}$}

\newcommand{\msini}{$M_{P} \sin i~$}

\lefthead{Arriagada {\it et~al.\/}}
\righthead{New Magellan Planets}
\begin{document}

\title{ Five Long-period Extrasolar Planets in Eccentric orbits from the Magellan Planet Search Program\altaffilmark{1}}

\author{Pamela Arriagada\altaffilmark{2},
R. Paul Butler\altaffilmark{3}, 
Dante Minniti\altaffilmark{2}, 
Mercedes L\'opez-Morales\altaffilmark{3,6}, 
Stephen A. Shectman\altaffilmark{4},
Fred C. Adams\altaffilmark{5},
Alan P. Boss\altaffilmark{3},
John E. Chambers\altaffilmark{3}}

\authoremail{parriaga@astro.puc.cl}

\altaffiltext{1}{Based on observations obtained with
the Magellan Telescopes, operated by the Carnegie
Institution, Harvard University, University of Michigan,
University of Arizona, and the Massachusetts Institute
of Technology.} 

\altaffiltext{2}{Department of Astronomy, Pontificia
Universidad Cat\'olica de Chile, Casilla 306, Santiago 22, Chile}

\altaffiltext{3}{Department of Terrestrial Magnetism, Carnegie
of Washington, 5241 Broad Branch Road NW,
Washington D.C. USA 20015-1305}

\altaffiltext{4}{Carnegie Observatories,
813 Santa Barbara Street, Pasadena, CA USA 91101}

\altaffiltext{5}{Astronomy Department, University of Michigan,
Ann Arbor, MI USA 48109}

\altaffiltext{6}{Hubble Fellow}

\begin{abstract}

Five new planets orbiting G and K dwarfs have emerged from
the Magellan velocity survey.
These companions are jovian-mass planets in eccentric 
($e\geq0.24$) intermediate and long-period orbits. 
HD 86226b orbits a solar metallicity G2 dwarf.  
The \msini mass of the planet is 1.5 \mjup, the semi-major axis is 2.6 AU, and the eccentricity 0.73.  HD 129445b orbits a metal rich G6 dwarf.  The minimum mass of the planet is \msini=1.6 \mjupe, the semi-major axis is 2.9 AU, and the eccentricity 0.70.  HD 164604b orbits a K2 dwarf.  The \msini mass is 2.7 \mjupe, semi-major axis is 1.3 AU, and the eccentricity is 0.24. HD 175167b orbits a metal rich G5 star.  The  \msini mass is 7.8 \mjupe, the semi-major axis is 2.4 AU, and the eccentricity 0.54.  HD 152079b orbits a G6 dwarf.  The
\msini mass of the planet is 3 \mjupe, the semi-major axis is 3.2 AU, and the eccentricity is 0.60.
\end{abstract}


\section{Introduction}
\label{intro}

During the past fifteen years, Doppler velocity surveys have uncovered more than 350 extra solar planets around late F, G, K, and M stars within 100 parsecs. This planetary sample covers a wide variety of masses, orbital periods and eccentricites (Butler et al. 2006, Udry \& Santos 2007). Most of these planets are jovian-mass with semimajor axes less than 2 AU.
Recent discoveries include Neptune-mass and terrestrial-mass planets with orbital periods of days to weeks (Rivera et al. 2005; Udry et al. 2006; Mayor et al. 2009; Vogt et al. 2009; Rivera et al. 2009), and solar system analogs with periods $\ge$ 10 years (Jones et al. 2009; Marcy et al. 2002).
While Doppler velocity surveys are increasingly oriented towards finding terrestrial mass planets in small orbits, intermediate and long period companions around nearby stars continue to emerge, and are the primary targets for next generation imaging and interferometric missions.

Since planet formation and evolution theories were in the past based on
our solar system, most planetary systems were expected to have circular or
low eccentricity orbits. Instead the observed range of exoplanet
eccentricities ranges from 0 to 0.93, with a median of e =0.24. The origin
of exoplanet eccentricities remains as a basic, unanswered question for
planet formation and evolution theory. Planets are believed to form on
roughly circular orbits, necessitating a mechanism for pumping up their
orbital eccentricities. Possible mechanisms include gravitational
scattering by close encounters with other planets on crossing orbits
(e.g., Weidenschilling \& Marzari 1996; Rasio \& Ford 1996), the Kozai
(1962) mechanism, where orbital eccentricities and orbital inclinations
can be interchanged in an oscillatory manner, and perturbations by other
stars (Malmberg \& Davies 2009). Disk torques during planet migration have
also been advanced, though the eccentricity enhancements obtained are
modest at best (e.g., Boss 2005; D'Angelo, Lubow, \& Bate 2006; Moorhead
\& Adams 2008). Recently, the Rossiter-McLaughlin effect has been used to
determine high orbital inclinations in highly eccentric planets around binary systems (Winn et al. 2009a) as well as possible retrograde orbits (Winn et al. 2009b, Narita et al. 2009, Anderson et al. 2009).
 %that at least three exoplanets discovered by Doppler
%spectroscopy have orbits that are inclined much more than the planets of
%our solar system (Winn et al. 2009b), 
% Both the inclination and
%eccentricity of the exoplanet HD 80606b, e.g., are very large (Winn et al.
%2009a). Given that HD 80606 has
%a stellar companion, HD 80607, it is unclear if the orbit of HD 80606b is
%a result of the Kozai mechanism, stellar perturbations, interactions with
%another planet, or all three. The discovery of the possibly retrograde
%orbit of the exoplanet HAT-P-7b (Winn et al. 2009b) implies that dynamical
%interactions can have a major effect on exoplanet orbits. 
Understanding which dynamical interactions are responsible for these orbital
peculiarities will require completing the census of exoplanet
eccentricities and inclinations.

In this paper we report the discovery of five eccentric Jupiter-mass
planets from the Magellan Planet Search Program. To date the Magellan
program has discovered 11 extra-solar planets, including the five reported
here (Lopez-Morales et al. 2008; Minniti et al. 2009).

\section{The Magellan Planet Search Program}
\label{obs_01}

The Magellan Planet Search Program began taking data in Dec 2002 using the MIKE echelle spectrograph (Bernstein et al. 2003), mounted on the 6.5-m Magellan II (Clay) telescope located at Las Campanas Observatory in Chile.
Using a 0.35 arc-sec slit, MIKE obtains spectra with  a resolution of R $\sim$ 50000, covering the wavelength range from 3900--6200 \AA~ divided into a red and a blue CCD. 
An Iodine absorption cell (Marcy \& Butler 1992) is mounted in
front of the MIKE entrance slit, imprinting the reference Iodine
spectrum directly on the incident starlight, providing a
wavelength scale and a proxy for the spectrometer point-spread-function (Butler et al. 1996).  The Iodine cell is a temperature
controlled sealed pyrex tube, such that the column density of Iodine remains constant indefinitely.

The Iodine spectrum (5000 - 6200 \AA) falls
on the red CCD.  The blue CCD captures the CaII H and K lines
used to monitor stellar activity.
We have monitored a number of stable main sequence stars with
spectral types ranging from late F to mid K.  Examples of these
are shown in Figures 1 and 2 of Minniti et al. 2009.  As these
figures demonstrate, the Magellan/MIKE system currently achieves
measurement precision of 5 \ms.  The internal measurement
uncertainty of our observations is typically 2 to 4 \ms, suggesting
the Magellan/MIKE system suffers from systematic errors at the
3 to 4 \ms level.  To account for this the velocity uncertainties
reported in this paper have 3 \ms is added in quadrature
to the internally derived uncertainties. 

The Magellan planet search program is surveying $\sim$400
stars ranging from F7 to M5.  A histogram of the $B-V$ colors
of the Magellan planet search stars is shown in Figure 1.
Stars earlier than F7  
do not contain enough Doppler information to achieve precision
of 5 \ms, while stars later than M5 are too faint even for a 6.5-m
telescope.  The stars in the Magellan program have been chosen to minimize
overlap with the AAT 3.9-m and Keck 10-m surveys.  
Subgiants have not been removed.  Stellar jitter for
subgiants is small, $\lesssim$ 5 m/s (Johnson et al. 2007).
Stars more than 2 magnitudes above the main sequence have
much larger jitter, thus have been removed from the observing
list based on Hipparcos distances (Perryman et al. 1997, ESA 1997).

Stars with known stellar companions within 2 arcsec are also
removed from the observing list as it is operationally difficult
to get an uncontaminated spectrum of a star with a nearby
companion.  Otherwise there is no bias against observing
multiple stars.  The Magellan target stars also contain no bias
against brown dwarf companions or against metallicity.

\section{High-eccentricity Jupiter-mass planets from the Magellan Survey}
\label{obs_02}

This paper reports the discovery of five new planet--mass
candidates.  The stellar properties of
the host stars are given in Table 1.  The first two columns
provide the HD catalog number and  the Hipparcos catalog number
respectively. 
The stellar masses are taken from Allende Prieto et al. (1999), [Fe/H] are taken from Holmberg et al. (2007, 2009). Spectral types are taken from the Simbad database.  

Figure 2 shows the H line for the 5 stars reported in this paper, 
in ascending order of B-V.  The Sun (bottom) is shown for
comparison.  %Mt. Wilson S values (Baliunas et al. 1995) have
%been measured from the H \& K lines in the Magellan spectra.
%The log(R'$_{\rm HK}$) values are listed in Table 1.
Four of these stars are chromospherically quiet.  The only
star showing activity is the K2 dwarf HD 164604.
Active K dwarfs have significantly lower radial-velocity
``jitter'' than F or G stars (Santos et al. 2000;
Wright 2005).
The expected photospheric radial velocity jitter  for all
five of these stars is $< 3$ m/s.

The best-fit orbital parameters of the companions are listed
in Table 2.  These are all massive planets with large signals  
(K $>$ 35 \ms).  Due to the sparseness of some of these data
sets, the semiamplitudes are poorly constrained.  The uncertainties
in the orbital parameters are calculated via a Monte Carlo approach
as described in Marcy et al. (2005).  The individual Magellan Doppler
velocity measurements are listed in Tables 3 through 5.  The properties
of the host stars and of their companions are discussed in turn below.

\subsection{HD 164604}

HD 164604 is a K2 V dwarf with $V=9.7$ and $B-V=1.39$. 
The \emph{Hipparcos} parallax (Perryman et al. 1997)
gives a distance of 38.46 pc and an absolute visual magnitude
$M_V= 6.72$. %The star is chromospherically quiet, with log(R'$_{\rm HK}$)$=-$5.13. 
Its metallicity is [Fe/H]$=-0.18$ (Holmberg et al. 2009).

Eighteen Magellan Doppler velocity observations of HD 164604
spanning 6 years have been made, as shown in Figure 3
and listed in Table 3.  The observations span three full orbital
periods.  The period of the best-fit Keplerian orbit
is $P=1.66$ years,
the semi-amplitude is $K=77$ \ms, and the eccentricity
is $e=0.24\pm0.14$. The RMS
of the velocity residuals to the Keplerian fit is 7.50 \ms.
The reduced $\chi_{\nu}$ of the Keplerian fit is 2.7.
Assuming a typical mass for a K2V star of \mstar=0.8 \msun, the minimum
mass of the companion is \msini=2.7 \mjupe, and the orbital
semi-major axis is 1.3 AU.

\subsection{HD 129445}

HD 129445 is a G6 V star with $V=8.8$ and $B-V=0.756$. The \emph{Hipparcos}
parallax (Perryman et al. 1997) gives a distance of  67.61 pc and
an absolute visual magnitude, $M_V=4.65$.  
%HD 48265 is chromospherically quiet, with log(R'$_{\rm HK}$)=-4.93.
Its metallicity is [Fe/H]$=0.25$ (Holmberg et al. 2009).

Seventeen Magellan Doppler velocity observations of HD 129445
have been obtained, as shown in Figure 4 and listed in Table 4.
The observations span a full orbital
period.  The semi-amplitude of the best-fit Keplerian orbit
is $K=38$ \ms, the period is $P=5.04$ years and the eccentricity is $e=0.70\pm0.10$. 
The RMS of the velocity residuals to the Keplerian orbital fit is 7.30 \ms.
The reduced $\chi_{\nu}$ of the Keplerian orbital fit is 2.5.
Assuming a stellar mass of \mstar=0.99 \msun (Allende Prieto et al. 1999) we derive a minimum
mass of \msini=1.6 \mjup and an orbital semi-major axis of 2.9 AU.

\subsection{HD 86226}
HD 86226 is a G2 V star with $V=7.93$ and $B-V=0.64$. The \emph{Hipparcos}
parallax (Perryman et al. 1997) gives a distance of  42.5 pc and
an absolute visual magnitude, $M_V=4.78$.
%HD 28185 is chromospherically quiet with log(R'$_{\rm HK}$)=-4.81.
Its metallicity is [Fe/H]$=-0.04$ (Holmberg et al. 2009).

Thirteen Magellan Doppler velocity observations have been
made of HD 86226 over 6.5 years, as shown in Figure 5 and listed
in Table 5.  These observations span a full orbital period.  
The best-fit Keplerian orbit to the Magellan data yields a period
$P =4.20$ years, a semi-amplitude ($K$) of 37 \ms, and an eccentricity
$e = 0.73\pm0.21$.
The RMS of the velocity residuals to the Keplerian orbital fit is 6.27 \ms.
The reduced $\chi_{\nu}$ of the Keplerian orbital fit is 1.82.
Given the stellar mass \mstar=1.02 \msun (Allende Prieto et al. 1999), the minimum mass
of the planet is \msini=1.5 \mjup with an orbital semi-major
axis of 2.6 AU.

\subsection{HD 175167}
HD 175167 is a G5 IV/V  star with $V=8.01$ and $B-V=0.75$. The \emph{Hipparcos}
parallax (Perryman et al. 1997) gives a distance of  67.02 pc and
an absolute visual magnitude, $M_V=3.88$, consistent with early
evolution off the main sequence.
%HD 28185 is chromospherically quiet with log(R'$_{\rm HK}$)=-4.81.
Its metallicity is [Fe/H]$=0.19$ (Holmberg et al. 2009).

Thirteen Magellan Doppler velocity observations have been
made of HD 175167 spanning 5 years, as shown in Figure 6 and listed
in Table 6.  These observations span a full orbital period.  
The best-fit Keplerian orbit to the Magellan data yields a period
$P = 3.43 $ years, a semi-amplitude ($K$) of 161 \ms, and an eccentricity
$e = 0.54\pm0.09$.
The RMS of the velocity residuals to the Keplerian orbital fit is 6.91 \ms.
The reduced $\chi_{\nu}$ of the Keplerian orbital fit is 2.7.
Given the stellar mass \mstar =1.102 \msun (Allende Prieto et al. 1999), the minimum mass
of the planet is \msini=7.8 \mjup with an orbital semi-major
axis of 2.4 AU.

\subsection{HD 152079}
HD 152079 is a G6 dwarf with $V=9.18$ and $B-V=0.71$. The \emph{Hipparcos}
parallax (Perryman et al. 1997) gives a distance of  85.17 pc and
an absolute visual magnitude, $M_V=4.53$.
Its metallicity is [Fe/H]$=0.16$ (Holmberg et al. 2009).

Fifteen Magellan Doppler velocity observations have been
made of HD 152079 over 5.7 years, as shown in Figure 7 and listed
in Table 7.  
The best-fit Keplerian to the Magellan data yields a period
$P = 5.7$ years, a semi-amplitude ($K$) of 58 \ms, and an eccentricity
$e = 0.60 \pm 0.24$.
The RMS of the velocity residuals to the Keplerian orbital fit is 3.58 \ms.
The reduced $\chi_{\nu}$ of the Keplerian orbital fit is 0.8.
Given the stellar mass \mstar=1.03 \msun (Allende Prieto et al. 1999), the minimum (\msini)
mass of the planet is \msini=3.0 $M_{Jup}$, with a semi-major
axis of 3.2 AU.

\section{Discussion}

This paper reports the detection of five companions using Magellan/MIKE that have not been previously published. These candidates are high-eccentricity long-period jovian mass and larger planets 
orbiting nearby G and K dwarfs with metallicities ranging from [Fe/H]=-0.18 to [Fe/H]=0.19.\\

%Figure 8 shows the period $versus$ year of discovery for the all known planets.  The planets from this paper are highlighted in blue.
%As the existing Doppler survey programs push into their second decade many more long period planets emerge, allowing us to compare the architecture of the Solar System to extrasolar planetary systems.\\

To date, there are 273 well characterized %in refereed papers
known extrasolar planets, which show a wide range of eccentricities, from circular to about $e=$0.9 with a median eccentricity of 0.24, contrary to what it was expected before the first exoplanets were discovered. Circular orbits in planets with P$<$20 days can be explained by tidal circularization or orbit decay at periastron, however, the origin of the observed eccentricity distribution is still under debate. Currently, the most compelling explanation for the observed high
eccentricities is that they result from planet-planet scattering
interactions within systems that contain multiple companions. These interactions presumably take place after the epoch of planet
formation, or perhaps during its latter stages. Scattering naturally
produces large eccentricities much like the observed distribution, and often results in the ejection of planets (e.g., Moorhead \& Adams 2005; Mazari 2005; Ford \& Rasio 2008; Juric \& Tremaine 2008;
Chatterjee et al. 2008). However, since planet-planet scattering alone
cannot explain the observed distribution of semimajor axes (scattering
cannot move planets far enough inward -- Adams \& Laughlin 2003),
migration due to disk torques is also likely to take place. These
disk torques can cause additional changes in eccentricity, including
excitation (Goldreich \& Sari 2003; Ogilivie \& Lubow 2004), damping (e.g., Nelson et al. 2000), or both (Moorhead \& Adams 2008). As a
result, a complete explanation for the observed eccentricity
distribution is still being constructed.

External bodies provide another source of perturbations that can
affect orbital eccentricity, even in systems that have reached
long-term stability.  Such action can be driven by implusive
perturbations from passing stars in the birth cluster, or more
gradually through distant stellar and/or massive planetary companions
(Kozai 1962; Holman et al. 1997; Mazeh et al. 1997; Zakamska \&
Tremaine 2004; Malmberg \& Davies 2009).  Simulations of two-body
interactions show how interactions between planets can lead to the
observed eccentricity distribution (see Juric \& Tremaine 2008 and the
aforementioned references). However, these simulations predict a
slightly larger number of very eccentric ($e > 0.5$) planets than the
observed distribution. On the other hand, Malmberg \& Davies (2009)
simulate planetary systems in binaries and study how the orbital
elements can be affected by perturbations exerted by the second
component; they find good agreement with the observed distribution of
eccentricities for extrasolar planets with semimajor axes between 1
and 6 AU.

Our newly discovered candidates span eccentricities from 0.24 to 0.73, and semi major axis from 1.3 to 3.2 AU.  
The parent stars of four of the candidates are not part of known binary systems and their RV curves show no other low mass companions. It is worth noting that in the period range $P>1000$ days, there are twelve planets with eccentricities higher than 0.5, as shown in figure 8. From these twelve planets, there is just one confirmed to be part of a binary system, and only three of them have eccentricities higher or similar to HD 129445, none of which belong to binary systems.  
Of the five planets reported in this paper, the lowest eccentricity value corresponds to HD 164604, the only candidate that shows a drift in velocity which indicates the presence of an additional outer body with an orbital period longer than 6 years.  In this case, the mechanism described by Malmberg \& Davies (2009) could explain the planet's eccentricity. It is also worth noting that this planet spends part of its orbit in the habitable zone of its parent star ($\sim 0.9$ AU).
Ongoing discoveries and further characterization of long period planets will lead to a better understanding of the origin of eccentric planet orbits.\\

To estimate the feasibility of performing an astrometric follow up of our candidates, we have calculated their astrometric amplitude (218, 132, 117, 470, 168 $\mu$arcseconds for  HD164604, HD129445, HD 86226, HD 175167 and HD 152079 respectively ). Ground-based surveys carried on CCD mosaic cameras mounted on medium-sized telescopes such as CAPSCam (Boss et al. 2009) can now achieve a precision of the order of milliarcsecond(mas), making it is beyond the reach of the astrometric signature of our planetary companions from the ground with current technology. Hipparcos (Perryman 2008) data, provide positions with a precision of 1 mas for fairly bright stars and 0.5 mas for some stars after refinement (van Leeuwen 2007), which is still too low to detect such small signatures. To date, two of these amplitudes could only be reached using HST observations (Benedict et al. 2002c, Benedict et al. 2008). In the future, however, optical space-based astrometric missions such as J-MAPS, Gaia, and SIM will make possible to reach $\mu$as precision, making plausible to observe such signature.

Imaging follow-up of our candidates with current ground-based 8-m class telescopes or HST would be just as unsuccessful. Due to the required magnitude contrast with the parent star, the minimum angular separation at which $\sim$5 \mjup planets can be detected around solar-type stars is greater than 0.4 arcsec (Neuh\"auser et al. 2005, Biller et al. 2007, Chun et al. 2008, Lagrange et al. 2009, Kasper et al. 2009), while these newly discovered planets, although long period, have angular separations of less than 0.1 arcsec, being too far to be reached by these instruments . They will be, however, main targets of next generation 30-m class telescopes equipped with Adaptive Optics and future interferometers.

These new planets clearly fit an emerging
pattern that there is a dearth of planets with semi-major axes of
less than $\sim$0.5 AU, as seen in figure 9. Presumably this is
a signature of migration timescale versus formation timescale as a
function of distance from the star, as suggested by Ida \& Lin (2004).

%The Magellan Planet Search Program has found so far 12 low mass companions (L\'opez-Morales et al. 2008, Minniti et al. 2009 and this work),  from which one of them is a brown dwarf. This gives a rough statistic of 1 brown dwarf in 10 low mass companions, which would be consistent with a mass function that follows dN/dm $\sim$ 1/m, assuming brown dwarfs to have a mass of 10\mjup.

%Since this paper finds five more Jovian planets and no brown dwarfs,
%we should add a comment about the `Brown Dwarf Desert'. Of course, in
%the last paper, we did find a brown dwarf, but the statistics continue
%to indicate that they are rare. For example, `our' Magellan search
%sample now has 12 companions (2 + 5 + 5) and one is a brown dwarf.
%Although the statistics are incredibly sparse, we get order of
%magnitude 1 brown dwarf in 10, which is roughly consistent with a mass
%function that goes like dN/dm \sim 1/m (where I am implicitly assuming
%that brown dwarfs are about 10 times the mass of a Jupiter).  Note
%that I am not advocating over-interpreting our stastistics, only that
%we can make a general comment about how these previously inferred
%trends continue to hold up. Hope that makes sense.

\acknowledgements
We are grateful to the NIST atomic spectroscopy staff,
in particular to Dr. Gillian Nave and Dr. Craig Sansonetti, for
their expert oversight in calibrating our Iodine cell with the
NIST FTS.
RPB gratefully acknowledges support from NASA OSS grant NNX07AR4OG.
M.L-M. acknowledges support
provided by NASA through Hubble Fellowship
grant HF-01210.01-A awarded by the STScI,
which is operated by the AURA, Inc. for NASA,
under contract NAS5-26555.
DM and PA are supported by the Basal CATA PFB-06, 
FONDAP Center for Astrophysics 15010003, and FONDECYT 1090213.
The referee, Dr. Michael Endl, made many helpful suggestions
that significantly improved this paper.
This paper has made use of the Simbad and NASA ADS data bases.

\clearpage

\begin{deluxetable}{rrrlllll}
\tablecaption{Stellar Properties}
\label{candid}
\tablewidth{0pt}
\tablehead{
\colhead{Star} & \colhead{Star} & \colhead{Spec} & \colhead{M$_{\rm Star}$} & \colhead{V} & \colhead{B-V} & \colhead{[Fe/H]} & \colhead{d} \\
\colhead{(HD)} & \colhead{(Hipp)} &\colhead{type} & \colhead{(M$_{\odot}$)} & \colhead{(mag)}  &   &  & (pc) } 
\startdata
  164604 &   88414 & K2 ~V & 0.8 & 9.7 & 1.39 &   --0.18& 38 \\
  129445 &   72203 & G6 ~V & 0.99 & 8.8 & 0.756 &  0.25 & 67.61\\
  86226 &  20723 & G2 ~V & 1.02 & 7.93 & 0.64& --0.04 & 42.48 \\
   175167 &  20723 &  G5~IV/V   & 1.102 & 8.01 & 0.751& 0.19 & 67.02 \\
  152079 &  20723 & G6 ~V & 1.023 & 9.18 &0.711& 0.16 & 85.17 \\
\enddata
\end{deluxetable}

\clearpage

\begin{deluxetable}{rlllllllll}
\rotate
\tablecaption{Orbital Parameters}
\label{candid}
\tablewidth{0pt}
\tablehead{
\colhead{Star}  & \colhead{Period} & \colhead{$K$} & \colhead{$e$} & \colhead{$\omega$} & \colhead{$T_0$} & \colhead{\msini} & \colhead{$a$} & \colhead{N$_{obs}$} & \colhead{RMS} 
\\
\colhead{(HD)} & \colhead{(days)} & \colhead{(\ms)} &\colhead{ } & \colhead{(degrees)} & \colhead{(JD-2450000)}  & \colhead{(\mjup)} & {(AU)} & \colhead{ } & \colhead{(\ms)}
} 
\startdata
164604\tablenotemark{a} & 606.4 $\pm$9 & 77 $\pm$32 & 0.24 $\pm$0.14 & 51 $\pm$23 & 52674 $\pm$80 & 2.7 $\pm$1.3 & 1.3 $\pm$0.05 & 18 & 7.50\\
129445 &  1840 $\pm$55 & 38 $\pm$6   & 0.70 $\pm$0.10 &  163 $\pm$15 & 53093 $\pm$50 & 1.6 $\pm$0.6 & 2.9 $\pm$0.2 & 17 & 7.30 \\
86226 & 1534 $\pm$280 &  37 $\pm$15 & 0.73 $\pm$0.21 & 58 $\pm$50 & 52240 $\pm$290 & 1.5 $\pm$1.0 & 2.6 $\pm$0.4 & 13 & 6.27\\
  175167 & 1290 $\pm$22 & 161 $\pm$55 & 0.54 $\pm$0.09 & 342 $\pm$9 & 53598 $\pm$48 & 7.8 $\pm$3.5 & 2.4 $\pm$0.05 & 13 & 6.91\\
  152079 & 2097 $\pm$930 & 58 $\pm$18 & 0.60 $\pm$0.24 & 325 $\pm$37 & 53193 $\pm$260 & 3.0 $\pm$2.0 & 3.2 $\pm$2.1 & 15 & 3.58\\
\enddata
\tablenotetext{a}{Additional Velocity Slope is -15.9 $\pm$2.9 \ms per yr.}
\end{deluxetable}

\clearpage

\begin{deluxetable}{rrr}
\tablecaption{Velocities for HD 164604}
\label{vel164604}
\tablewidth{0pt}
\tablehead{
JD & RV & error \\
(-2452000)   &  (m s$^{-1}$) & (m s$^{-1}$)
}
\startdata
   808.7659  &   -12.1  &  8.9 \\
   918.5317  &    -6.0  &  7.6 \\
  1130.9311  &    84.2  &  4.7 \\
  1540.7199  &   -34.7  &  3.8 \\
  2011.5059  &   -57.8  &  5.5 \\
  2013.5185  &   -63.0  &  4.8 \\
  2277.7380  &    -5.6  &  4.4 \\
  2299.6480  &     1.5  &  4.5 \\
  2300.6352  &    -9.1  &  4.5 \\
  2339.5686  &    21.3  &  4.8 \\
  2399.4832  &    55.2  &  4.6 \\
  2926.8545  &   -11.7  &  4.1 \\
  2963.8563  &    24.3  &  4.6 \\
  2965.8500  &    12.3  &  4.8 \\
  2993.7397  &    15.1  &  4.4 \\
  3001.7597  &    28.1  &  4.2 \\
  3017.7019  &    35.1  &  4.4 \\
  3019.7039  &    43.0  &  4.1 \\
\enddata
\end{deluxetable}

\clearpage

\begin{deluxetable}{rrr}
\tablecaption{Velocities for HD 129445}
\label{vel 129445}
\tablewidth{0pt}
\tablehead{
JD & RV & error \\
(-2452000)   &  (m s$^{-1}$) & (m s$^{-1}$)
}
\startdata
   864.5311  &    26.1  &  8.2 \\
  1042.8730  &   -15.2  &  8.7 \\
  1127.8240  &   -35.8  &  4.2 \\
  1480.8541  &    15.7  &  5.7 \\
  1574.5786  &    22.5  &  4.8 \\
  1575.5511  &    29.4  &  4.4 \\
  1872.6777  &    43.7  &  4.2 \\
  2217.7257  &    33.3  &  4.3 \\
  2277.5928  &    37.5  &  4.8 \\
  2299.4993  &    31.6  &  4.3 \\
  2501.8506  &    44.2  &  4.1 \\
  2522.8417  &    32.9  &  4.4 \\
  2925.8091  &   -31.9  &  3.9 \\
  2963.7305  &   -40.3  &  4.2 \\
  2993.6537  &   -10.8  &  4.0 \\
  3001.6458  &   -21.4  &  4.2 \\
  3017.6200  &   -14.4  &  3.9 \\
\enddata
\end{deluxetable}

\clearpage

\begin{deluxetable}{rrr}
\tablecaption{Velocities for HD 86226}
\label{vel86226}
\tablewidth{0pt}
\tablehead{
JD & RV & error \\
(-2452000)   &  (m s$^{-1}$) & (m s$^{-1}$)
}
\startdata
   626.8679  &   -24.6  &  7.5 \\
   663.7551  &   -11.1  &  5.1 \\
  1041.6735  &   -12.4  &  6.9 \\
  1128.5597  &     2.4  &  4.2 \\
  1455.6305  &    11.5  &  4.6 \\
  1784.7926  &    20.4  &  4.8 \\
  2583.6051  &    -4.7  &  4.2 \\
  2843.8112  &    -1.5  &  6.3 \\
  2925.6391  &    11.9  &  4.2 \\
  2963.5403  &    11.3  &  4.0 \\
  2994.5061  &    11.9  &  4.1 \\
  3001.4805  &     9.4  &  4.5 \\
  3019.4585  &    19.6  &  4.2 \\
\enddata
\end{deluxetable}

\clearpage

\begin{deluxetable}{rrr}
\tablecaption{Velocities for HD 175167}
\label{vel175167}
\tablewidth{0pt}
\tablehead{
JD & RV & error \\
(-2453000)   &  (m s$^{-1}$) & (m s$^{-1}$)
}
\startdata
   189.7359  &  -140.3  &  4.2 \\
   190.7340  &  -138.4  &  4.3 \\
   191.7523  &  -135.6  &  4.4 \\
   254.5236  &  -124.5  &  4.2 \\
   654.5074  &   146.1  &  4.3 \\
   656.5134  &   152.6  &  4.0 \\
  1217.9281  &  -124.1  &  4.5 \\
  1339.6070  &  -137.3  &  4.3 \\
  1725.6182  &   -73.1  &  3.8 \\
  1965.8675  &   121.6  &  4.2 \\
  1993.7656  &    72.3  &  4.1 \\
  2001.7759  &    81.9  &  4.3 \\
  2017.7389  &    49.1  &  3.9 \\
\enddata
\end{deluxetable}

\clearpage

\begin{deluxetable}{rrr}
\tablecaption{Velocities for HD 152079}
\label{vel152079}
\tablewidth{0pt}
\tablehead{
JD & RV & error \\
(-2452000)   &  (m s$^{-1}$) & (m s$^{-1}$)
}
\startdata
  917.4972  &   -24.3  &  6.2 \\
  1542.6649  &    22.5  &  3.3 \\
  1872.8022  &    -8.5  &  2.5 \\
  1987.5436  &   -10.3  &  2.8 \\
  1988.5202  &   -12.6  &  2.7 \\
  2190.8274  &   -13.7  &  2.9 \\
  2277.6950  &   -19.7  &  3.4 \\
  2299.6134  &   -19.6  &  3.3 \\
  2725.5353  &   -35.1  &  2.6 \\
  2925.9161  &   -29.2  &  2.4 \\
  2963.7753  &   -22.6  &  2.7 \\
  2993.7093  &   -27.5  &  2.4 \\
  3001.7291  &   -25.3  &  2.9 \\
  3017.6624  &   -28.5  &  2.4 \\
  3019.6938  &   -22.5  &  2.2 \\
\enddata
\end{deluxetable}

\clearpage

\begin{figure}
\includegraphics[angle=90,width=\textwidth]{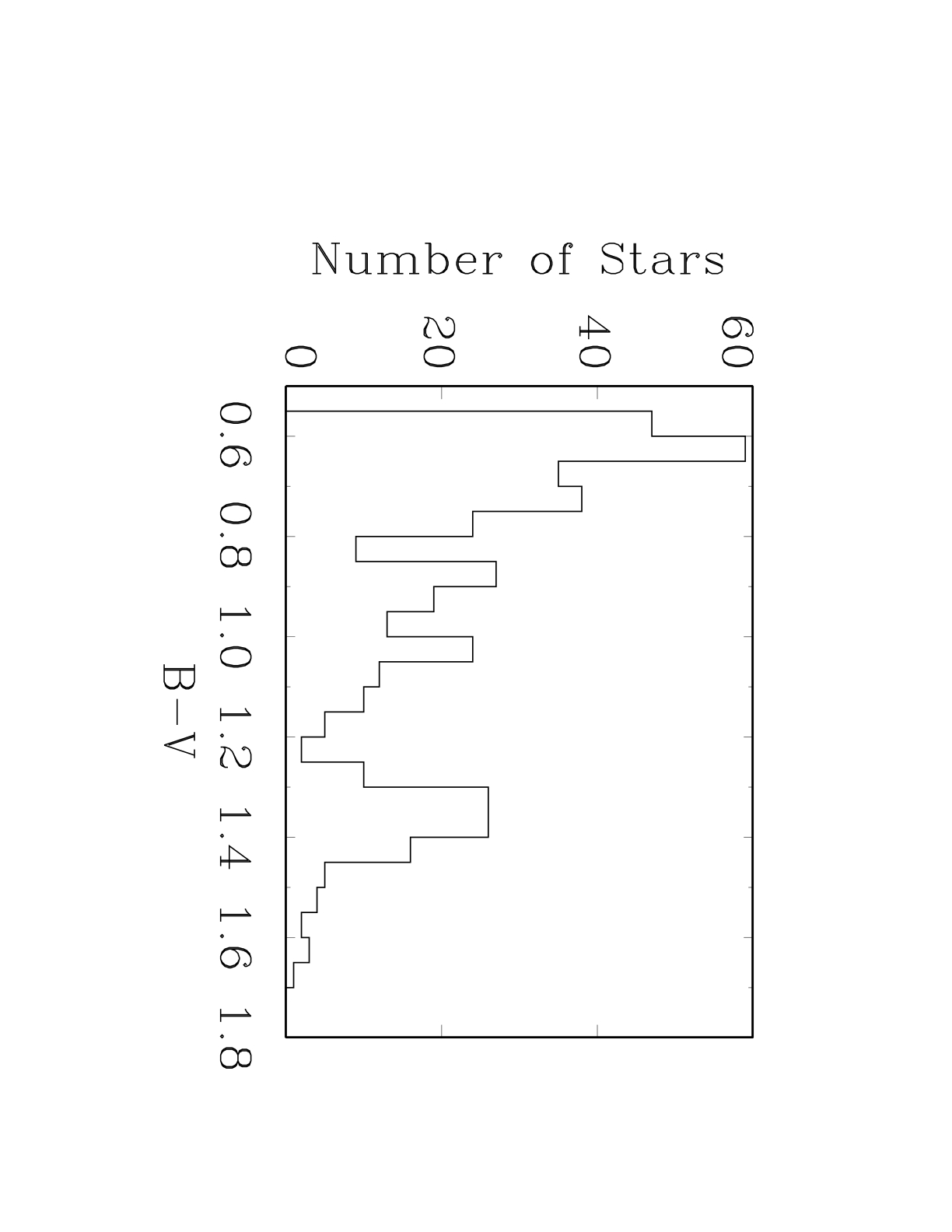}
\caption{B-V histogram of Magellan Planet Search Stars.
The distribution peaks around sun-like stars and
diminishes for later spectral types.  There is a
secondary peak in the distribution around B-V = 1.35,
reflecting our bias toward adding the nearest M dwarfs. }
\label{fig1}
\end{figure}

\begin{figure}
\includegraphics[width=\textwidth]{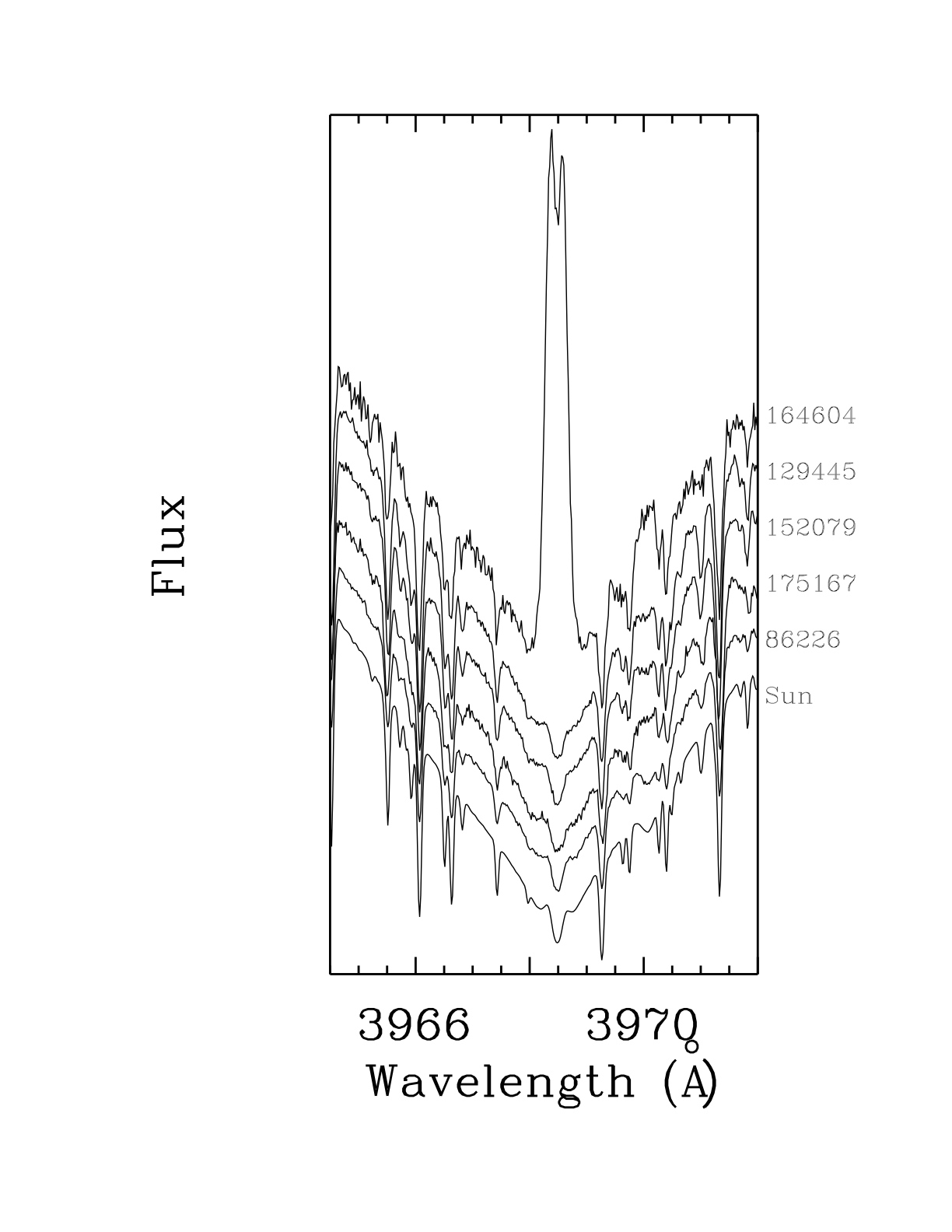}
\caption{Ca II H line cores for the five target G dwarfs in ascending order of $B - V$.The HD catalog number of each star is shown along the right edge. The Sun is shown for comparison. }
\label{fig2}
\end{figure}

\begin{figure}
\includegraphics[angle=90,width=\textwidth]{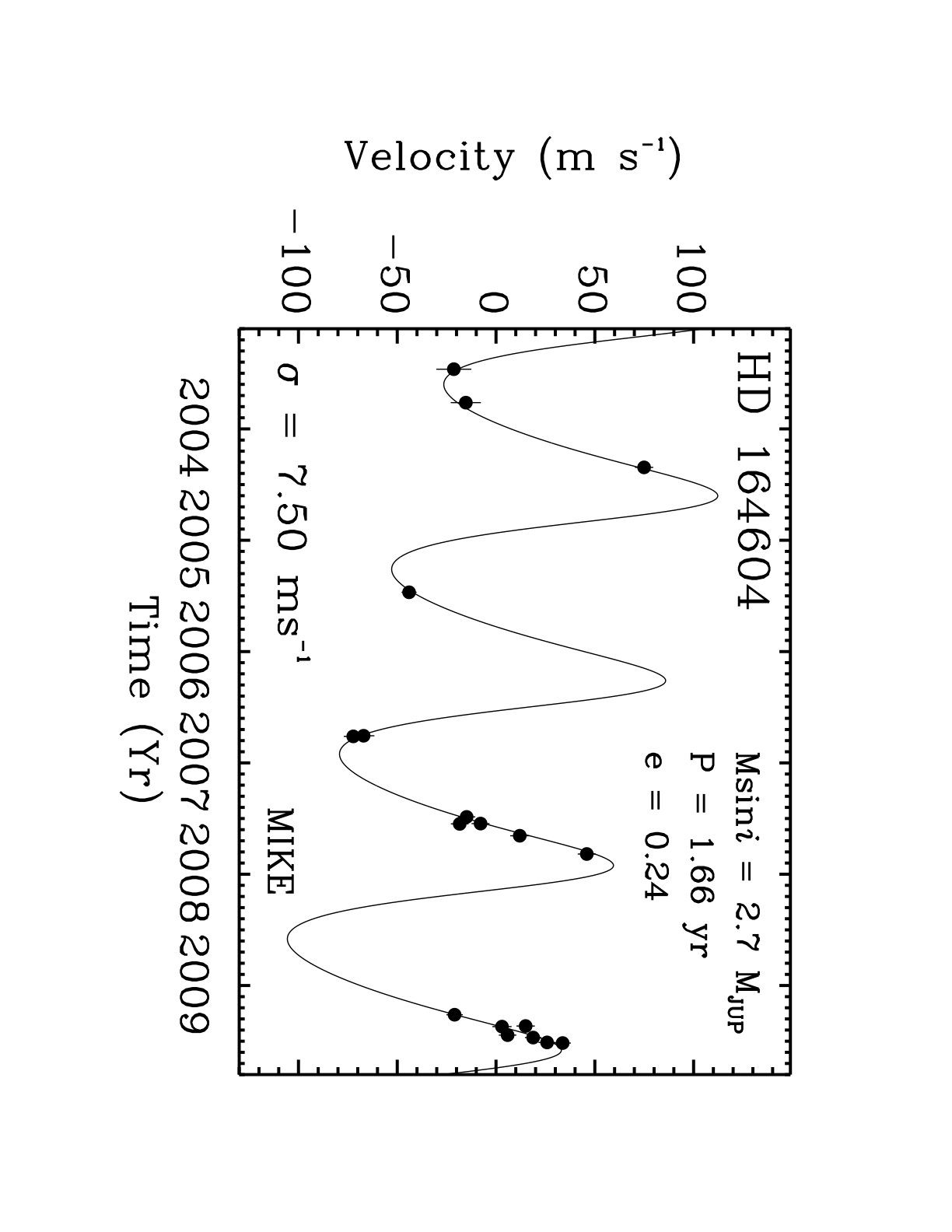}
\caption{Doppler velocities for HD 164604 (K2 V).
The solid line is a Keplerian orbital fit with a
period of 1.66 years, a semi-amplitude of 77.4 \ms,
and an eccentricity of 0.24, yielding a minimum
companion mass (\msini) of 2.7 \mjup \.  The
RMS of the Keplerian fit is 7.50 \ms.  An additional
linear trend of -15.9 \ms per year provides evidence
for a massive outer companion with a period greater than
7 years and a semiamplitude greater than 50 \ms.}
\label{fig3}
\end{figure}

\begin{figure}
\includegraphics[angle=90,width=\textwidth]{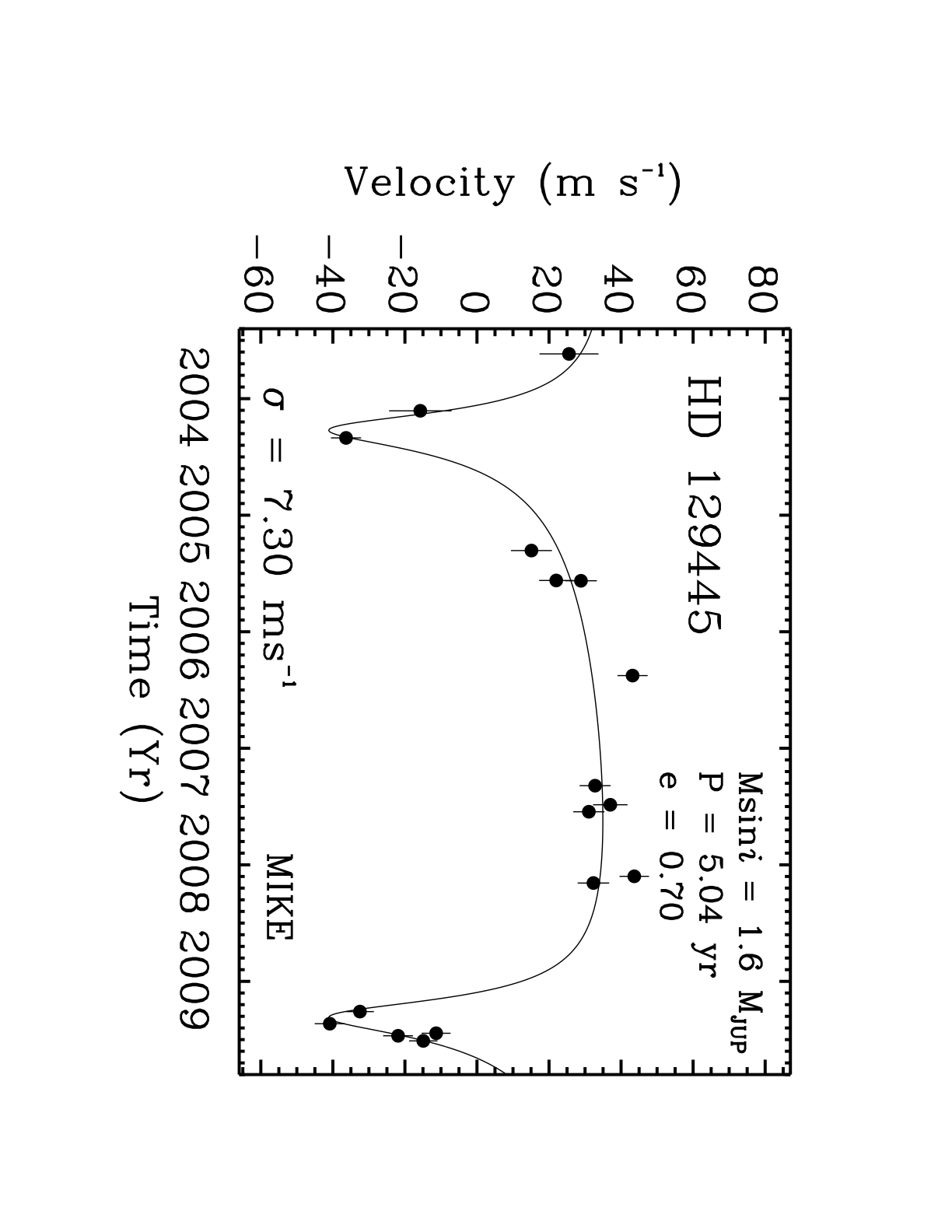}
\caption{Doppler velocities for HD 129445 (G6 V).
The solid line is a Keplerian orbital fit with a
period of 5.04 years, a semi-amplitude of 38 \ms,
and an eccentricity of 0.70, yielding a minimum
(\msini) companion mass of 1.6 \mjup \.  The
RMS of the Keplerian fit is 7.30 \ms.}
\label{fig4}
\end{figure}

\begin{figure}
\includegraphics[angle=90,width=\textwidth]{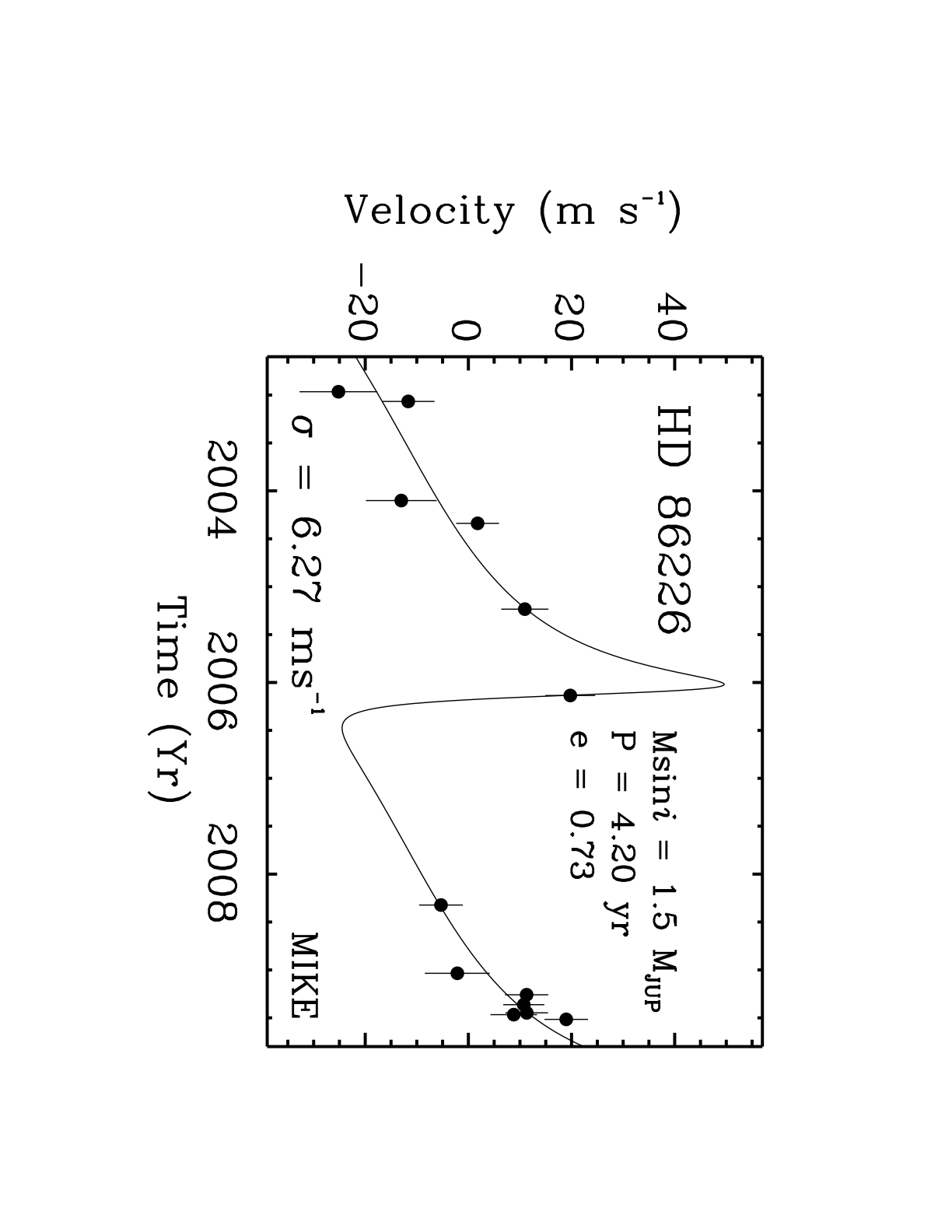}
\caption{ 
Doppler velocities for HD 86226 (G2 V).
The solid line is a Keplerian orbital fit with a
period of 4.20 years, a semi-amplitude of 37 \ms,
and an eccentricity of 0.73, yielding a minimum
(\msini) of 1.5 \mjup \ for the companion.  The
RMS of the Keplerian fit is 6.27 \ms.}
\label{fig5}
\end{figure}

\begin{figure}
\includegraphics[angle=90,width=\textwidth]{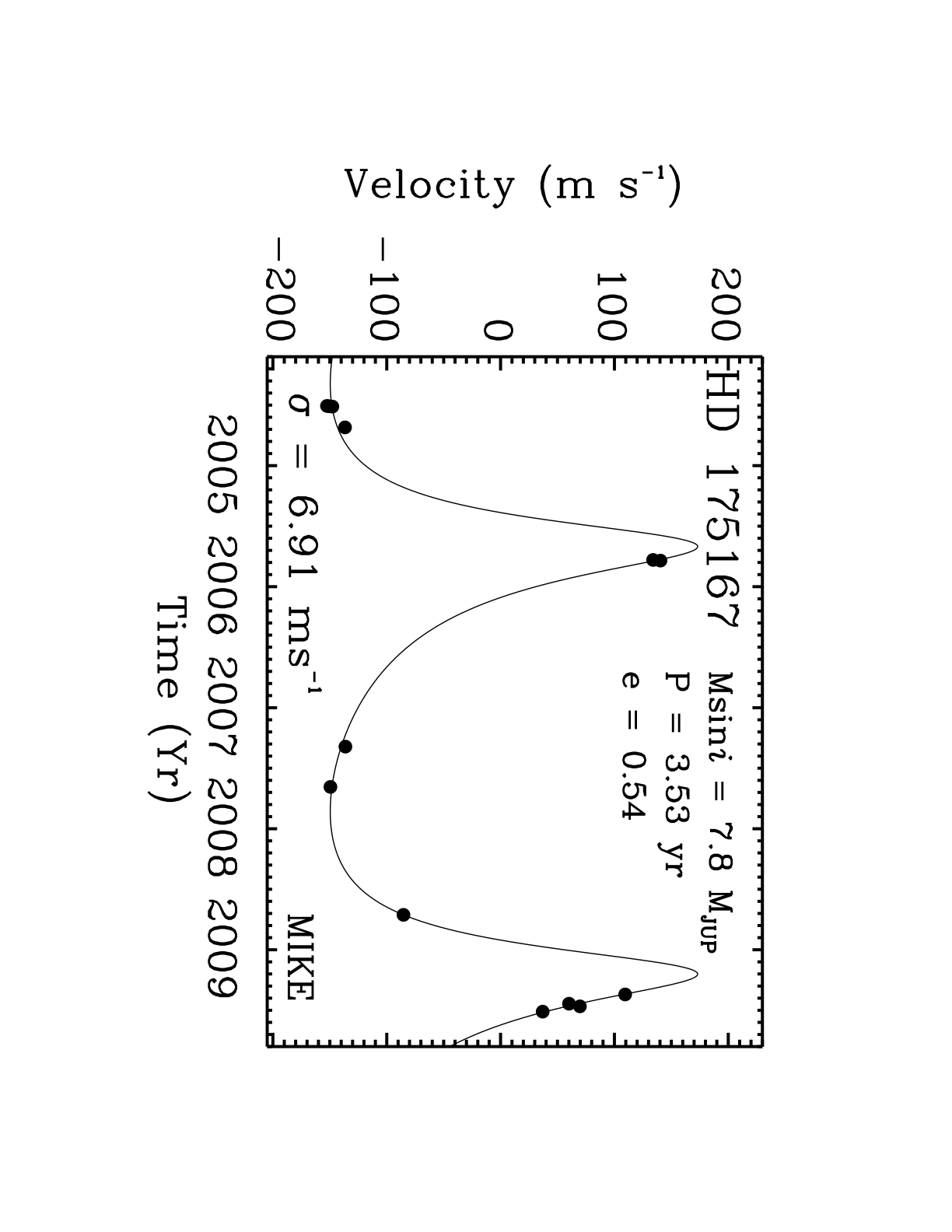}
\caption{ 
Doppler velocities for HD 175167 (G5 IV/ V).
The solid line is a Keplerian orbital fit with a
period of 3.53 years, a semi-amplitude of 161 \ms,
and an eccentricity of 0.54, yielding a minimum
(\msini) companion mass of 7.8 \mjup \.  The
RMS of the Keplerian fit is 6.91 \ms.}
\label{fig6}
\end{figure}

\begin{figure}
\includegraphics[angle=90,width=\textwidth]{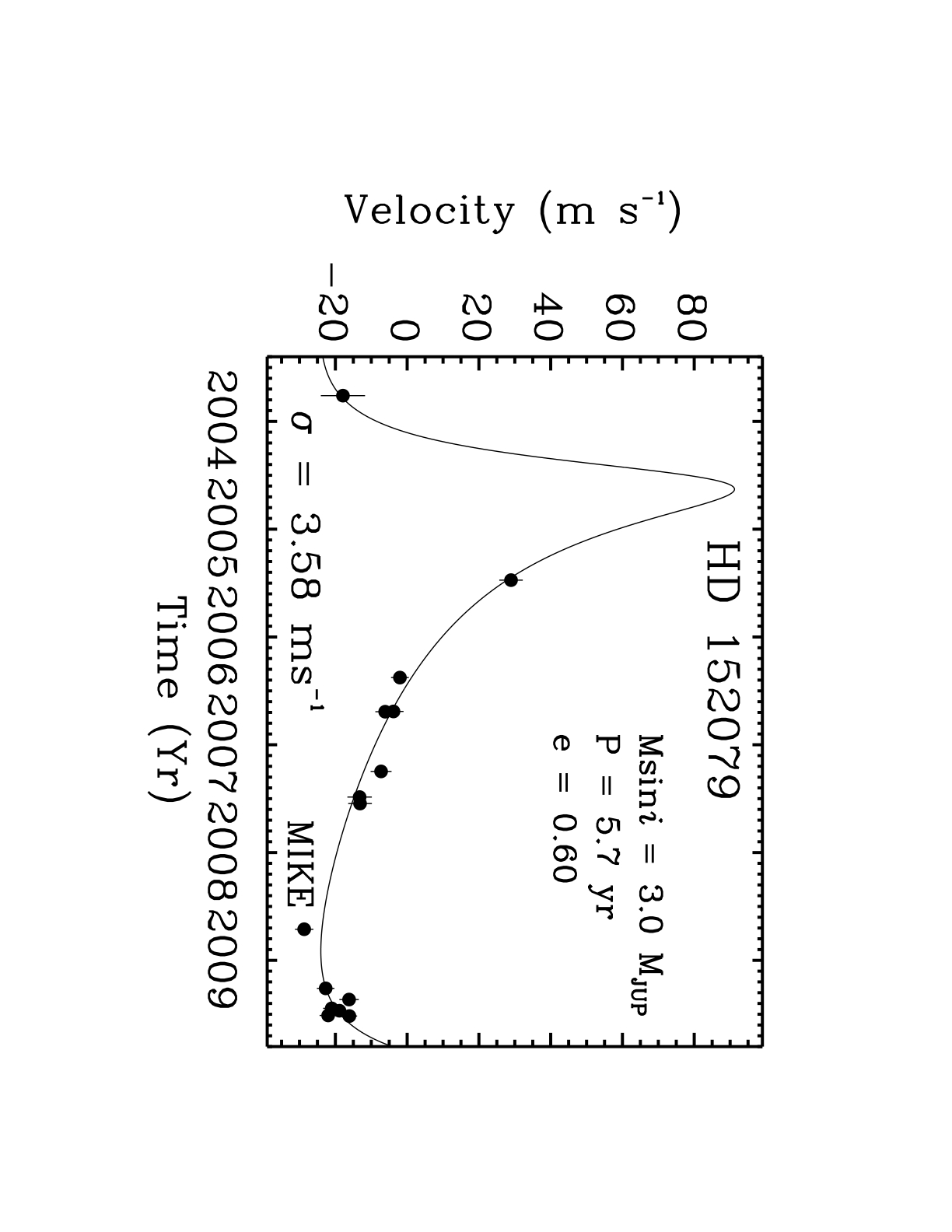}
\caption{ 
Doppler velocities for HD 152079 (G6 V).
The solid line is a Keplerian orbital fit with a
period of 5.04 years, a semi-amplitude of 33.1 \ms,
and an eccentricity of 0.56, yielding a minimum
(\msini) of 3.0 \mjup \ for the companion.  The
RMS of the Keplerian fit is 3.58 \ms.}
\label{fig7}
\end{figure}

%\begin{figure}
%\includegraphics[angle=00,width=\textwidth]{figure8.ps}
%\caption{Year of discovery vs. orbital period for known extra-solar planets. Our five newly detected planets are presented in blue. Circle sizes are proportional to the planet's mass.}
%\label{fig4}
%\end{figure}

\begin{figure}
\includegraphics[angle=00,width=\textwidth]{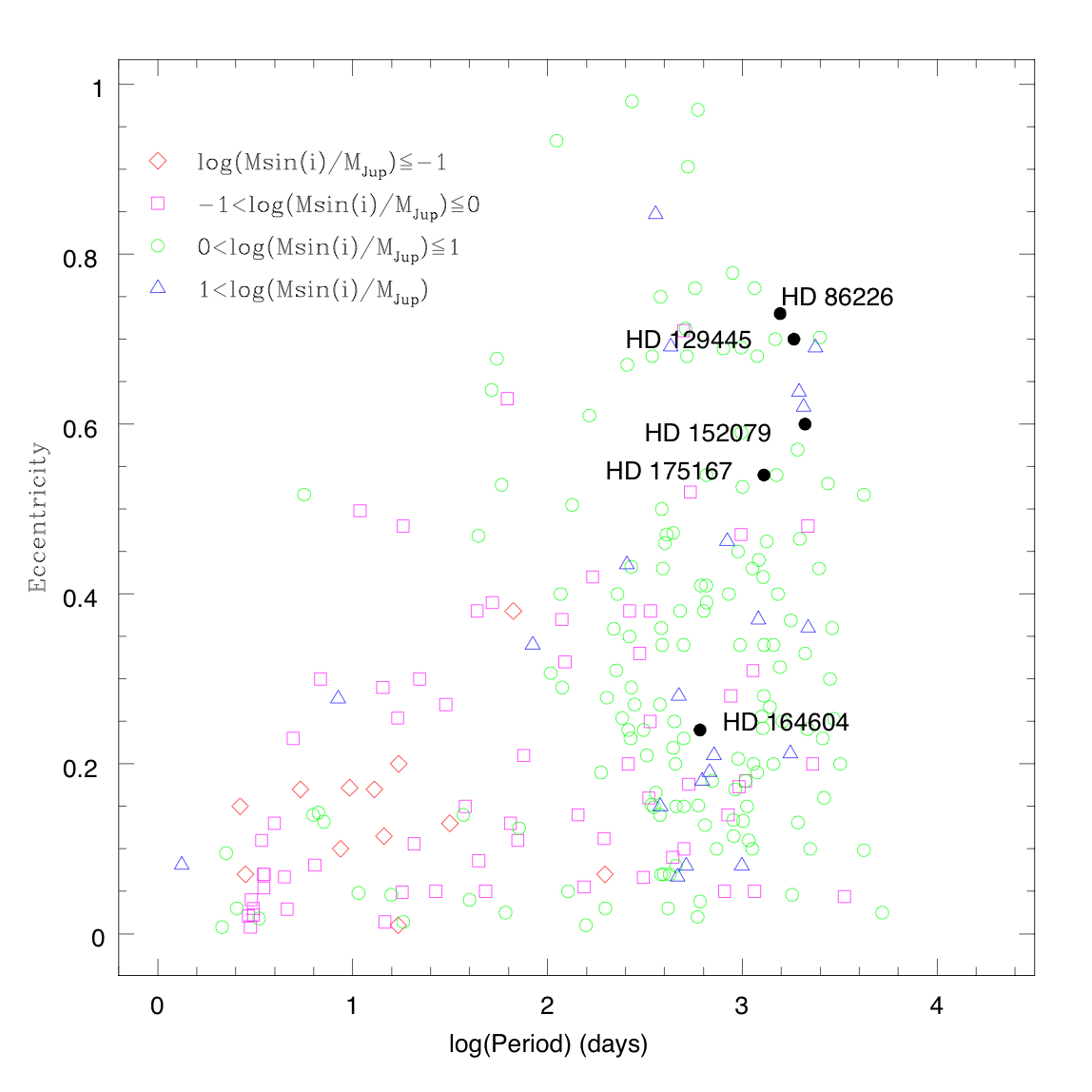}
%\caption{Eccentricities vs. orbital period of known extra-solar planets, where the planets reported in this paper are highlighted in blue. Circle sizes are proportional to the planet's mass. Note that three of the planets announced in this paper have eccentricities higher than 0.5. }
\caption{Eccentricities vs. orbital period of known extra-solar planets, where the planets reported in this paper are in filled symbols. Different symbols denote different mass ranges. Note that four of the planets announced in this paper have eccentricities higher than 0.5. }
\label{fig8}
\end{figure}

\begin{figure}
\includegraphics[angle=00,width=\textwidth]{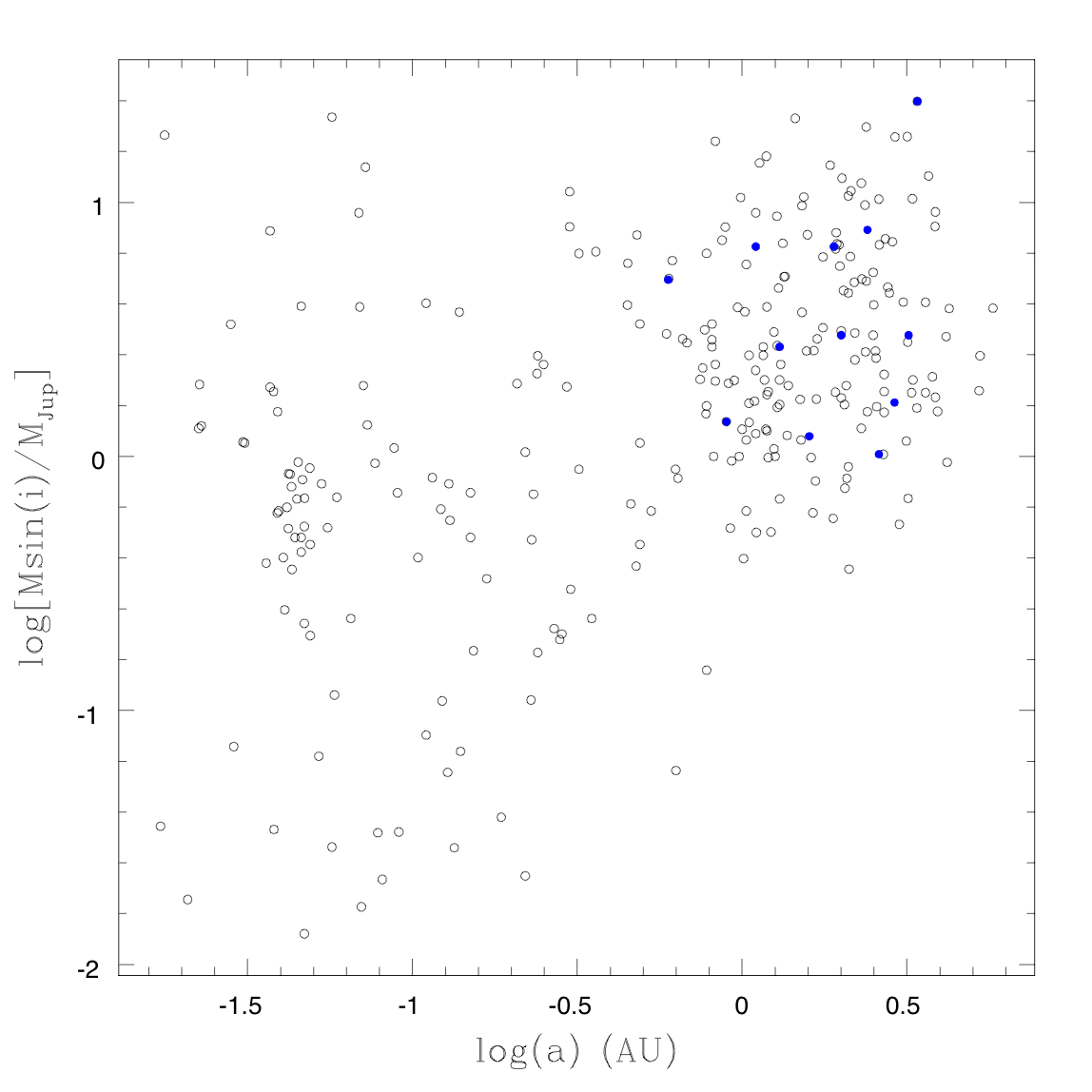}
\caption{Semimajor axis (a) versus Msini. All low-mass companions discovered by the Magellan Planet Search Program are highlighted as filled, blue circles. }
\label{fig9}
\end{figure}

\end{document}